\pdfoutput=1
\documentclass[fleqn,usenatbib]{mnras}
\usepackage{graphicx}
\usepackage{amsmath,amssymb}
\usepackage{lastpage}
\newcommand{\red}{\textcolor{black}}
\usepackage[all]{hypcap}

\usepackage{float}
\usepackage{subfigure}
\usepackage{afterpage}
\usepackage[usenames]{color}
\usepackage{yfonts}
\usepackage{mathrsfs}
\usepackage{upgreek}
\usepackage{hyperref}

\title{Reconstructing the gravitational field of the local universe}
\author[H.~Desmond, P.~G.~Ferreira, G.~Lavaux, J.~Jasche]{
Harry~Desmond$^{1,2,3}$\thanks{E-mail: harryd2@stanford.edu},
Pedro~G.~Ferreira$^1$,
Guilhem~Lavaux$^{4,5}$
and \newauthor{}
Jens~Jasche$^6$
\\
$^{1}$Astrophysics, University of Oxford, Denys Wilkinson Building, Keble Road, Oxford OX1 3RH, UK\\
$^{2}$Kavli Institute for Particle Astrophysics and Cosmology, Physics Department, Stanford University, Stanford, CA 94305, USA\\
$^{3}$SLAC National Accelerator Laboratory, Menlo Park, CA 94025, USA\\
$^{4}$Sorbonne Universit{\'é}s, UPMC Univ Paris 6 et CNRS, UMR 7095, Institut d'Astrophysique de Paris, 98 bis bd Arago, 75014 Paris, France\\
$^{5}$Sorbonne Universit{\'é}s, Institut Lagrange de Paris (ILP), 98 bis bd Arago, 75014 Paris, France\\
$^{6}$Excellence Cluster Universe, Technische Universit{\"ä}t M{\"ü}nchen, Boltzmannstrasse 2, D-85748 Garching, Germany
}

\pubyear{2017}

\begin{document}
\label{FirstPage}
\pagerange{\pageref{FirstPage}--\pageref{LastPage}}
\maketitle

\begin{abstract}
Tests of gravity at the galaxy scale are in their infancy. As a first step to systematically uncovering the gravitational significance of galaxies, we map three fundamental gravitational variables -- the Newtonian potential, acceleration and curvature -- over the galaxy environments of the local universe to a distance of approximately $200$ Mpc. Our method combines the contributions from galaxies in an all-sky redshift survey, halos from an N-body simulation hosting low-luminosity objects, and linear and quasi-linear modes of the density field. We use the ranges of these variables to determine the extent to which galaxies expand the scope of generic tests of gravity and are capable of constraining specific classes of model for which they have special significance. Finally, we investigate the improvements afforded by upcoming galaxy surveys.
\end{abstract}


\section{Introduction}
\label{sec:intro}

General Relativity (GR) has been the reigning paradigm of gravity for almost a century, and yet there is no shortage of alternatives. A range of possible reasons to extend or replace GR has been explored in the literature, including the attempt to overcome theoretical or conceptual difficulties, obviate the need for dark energy or alleviate the cosmological constant problem, account more precisely for astrophysical phenomena, or simply investigate the range of gravity theories consonant with basic physical principles or that may have interesting observational consequences (see e.g.~\citealt{ModGrav} and references therein). These diverse motivations have led to a large and heterogeneous parameter space of models which cannot be thoroughly probed by traditional tests. Instead, systematic progress will require the synthesis of evidence from across the range of scales accessible to observation and experiment.

To date, most effort has been devoted to probing gravity in three types of system: the laboratory, the Solar System, and the linear and quasi-linear cosmological regimes. In the laboratory, tests of the equivalence principle (EP) on which GR is premised have now reached $\mathcal{O}(10^{-13})$ precision~\citep{EP}, severely limiting deviations from the inverse square law. Within the Solar System, the Parameterised Post-Newtonian (PPN) framework has enabled the coefficients of general metric theories to be constrained at leading relativistic order, providing tight limits on the coupling of extra fields to matter and generic deviations from standard weak-field equations of motion~\citep{Nordtvedt,Will}. In cosmology, GR is combined with the hypotheses of dark matter and dark energy to form the $\Lambda$CDM model, which may be probed via its predictions for Big Bang Nucleosynthesis, the Cosmic Microwave Background, the expansion history of the universe and the growth rate of structure (e.g.~\citealt{Bull} and references therein). Analogously to PPN, frameworks have also been devised to test generic deviations from standard cosmological metrics (e.g.~\citealt{PPF}). In none of these cases has a convincing deviation from GR been found.

Nevertheless, these tests do not cover the full range of systems in which deviations from GR may appear, and modified gravity theories may be constructed which satisfy all current experimental bounds and yet exhibit divergent behaviour elsewhere. The possibility that GR may break in some systems -- or as a function of certain variables -- but not others decouples tests in different regimes, and introduces the possibility that novel gravitational signals may be sequestered in regions of parameter space so far unexplored. Indeed, the notion that physical theories break down at critical values of certain variables, and hence in select systems, is not new: Galilean Relativity gives way to Special at $v$ near $c$, and Newtonian gravity to GR at $\Phi$ near $c^2$. Other variables, in which present tests span only a limited range of values, may mark the onset of new gravitational regimes.

The aim of this paper is to contribute to the development of tests of GR in a qualitatively different and relatively under-explored regime: the galactic. In particular, we see the galaxy scale as able to extend the scope of gravitational probes in three main ways:

\begin{enumerate}

\item{} Many scalar-tensor theories of gravity must employ a ``screening mechanism'' to suppress their fifth force in the Solar System and laboratory. This may be achieved either by the scalar field acquiring a high mass and hence short range when large (chameleon screening;~\citealt{Chameleon}), or by kinetic (kinetic screening;~\citealt{kinetic}) or higher order (Vainshtein screening;~\citealt{Vainshtein}) terms in the Lagrangian becoming dominant. In each case the degree of screening is governed by a different function of the scalar field, which simulations have shown to correlate with simple Newtonian proxies: the potential $\Phi$ for chameleon screening, acceleration $\vec{a}$ for kinetic, and some function of the Riemann tensor -- we will call it the ``curvature'' $K$ -- for Vainshtein~\citep{Cabre,Khoury_LesHouches}. For typical scalar-tensor theories, the transitions between screened and unscreened regions occur at values of these proxies characteristic of galaxies and their environments~\citep{Hui, Jain_Vanderplas}; GR is recovered at the larger values that describe laboratory and Solar System tests. Further, for a broad class of chameleon-like theories laboratory fifth force constraints imply a range for the scalar field at cosmological densities of $\lesssim 1$ Mpc, rendering its impact on linear perturbations and the universe's expansion history negligible~\citep{Wang}. This makes the intermediate galaxy scale the ideal one at which to test such models.

\item{} A dependence of the kinematics of any system on the external gravitational field constitutes a violation of the equivalence principle (EP), and \textit{a fortiori} of GR. The EP has been tested precisely in the lab (at $a \approx 10 \: \text{m} \, \text{s}^{-2}$) and inner Solar System ($a \approx 10^{-2} \: \text{m} \, \text{s}^{-2}$). Galaxies, however, have characteristic accelerations at least six orders of magnitude lower, and hence allow for EP tests in a very different region of parameter space. Beside the case of screening described above (in which EP violation is typically effective at the macroscopic level, rather than fundamental at the level of the action), a paradigmatic example of a theory that exploits this gap to violate the EP is Modified Newtonian Dynamics (MOND;~\citealt{Milgrom1,Milgrom2,Milgrom3}). This phenomenon is known in MOND as the ``external field effect,'' whereby a large external acceleration can render a system's dynamics Newtonian or quasi-Newtonian even for internal accelerations much below the threshold value $a_0 \approx 10^{-10} \: \text{m} \, \text{s}^{-2}$. As explained further in Section~\ref{sec:modgrav}, galaxies' mass-to-light ratios and the shapes of their rotation and velocity dispersion profiles would be expected to correlate with external field strength when the latter was sufficiently high, enabling the theory to be tested by means of a prediction complementary to the more conventional ones concerning the internal field alone. Indeed, galaxy environments are among the only with accelerations at and around $a_0$, and hence are uniquely capable of testing this and related models.

\item{}~\citet{Baker} mapped out the theoretical and experimental gravity parameter space in terms of $\Phi$ and $K$, and discovered that the curvature values of galaxies are not currently probed by any observational test. These values separate the relatively well-studied small-scale regime at high $K$ from the `troubled' cosmological regime at low $K$, leading~\citet{Baker} to suggest they may be a natural place for novel gravitational physics to emerge. Since galaxies are the \emph{only} systems to inhabit this region of parameter space, they are the only ones \emph{in principle} capable of probing it, regardless of measurement precision.

\end{enumerate}

These motivations suggest we focus on three gravitational variables at the galaxy scale: the Newtonian potential $\Phi$, acceleration $\vec{a}$ and curvature $K$. More generally, these are among the most basic descriptors of the gravitational field, and may therefore be expected \textit{a priori} to characterise the transitions between gravitational regimes. In the weak-field limit for a set of point masses (as describes galaxies), they scale as $M/r$, $M/r^2$ and $M/r^3$ respectively, and hence provide a range of relative scalings with the variables $M$ and $r$ that determine an object's gravitational influence.

The first step towards identifying novel gravitational physics associated with one or more of these variables is to map out their values over the local universe. That is the task of this paper. In particular, we will build maps of $\Phi$, $\vec{a}$ and $K$ out to $\sim 200 \: \text{Mpc}$ by combining the contributions of galaxies measured in an all-sky survey, halos in an N-body simulation hosting galaxies too faint to be observed, and a smooth density field not captured by the halo model. Given a set of galaxies with measurements for potential modified gravity signals, these maps will allow the signals' correlations with each gravitational variable to be determined, and hence the theories above to be investigated and gaps in the experimental parameter space filled in. This will be the subject of future work.

The only previous work along these lines is~\citet{Cabre}, who focus on $\Phi$ as an estimator of the degree of screening in chameleon theories. Besides extending to $\vec{a}$ and $K$, our analysis will build on this pioneering study in several ways: we will use a more complete and homogeneous all-sky source catalogue, apply more sophisticated methods for determining source objects' mass distributions from their magnitudes, and calibrate and supplement the basic maps with N-body simulations and estimates of the quasi-linear density field respectively. Unlike~\citet{Cabre}, we will make no attempt to relate our proxies to features of specific theories through modified gravity simulations, but focus instead on simply mapping the gravitational variables as robustly as possible. This will allow our results to retain full generality for future application to any model in which these variables are significant.

The structure of this paper is as follows. In Section~\ref{sec:method} we describe our method for constructing maps of $\Phi$, $\vec{a}$ and $K$, and in Section~\ref{sec:results} we present the results. Section~\ref{sec:discussion} applies our findings to modified gravity, describes the main sources of uncertainty, and discusses the improvements in precision achievable by upcoming galaxy surveys. Section~\ref{sec:conclusion} concludes.

\section{Method}
\label{sec:method}

We make two general methodological comments before detailing our 3-step procedure.

While each of $\Phi$, $\vec{a}$ and $K$ has an ``internal'' and ``external'' component, the first due to an object's own mass and the second that of its environment, we will focus solely on the environmental contribution, which is a function purely of spatial position. The internal component depends on objects' masses, and must therefore be calculated specifically for a given test galaxy sample. As $\Phi$, $\vec{a}$ and $K$ are our estimators for the degrees of freedom in which GR may receive corrections, we refer to them hereafter as ``proxies.''

The Newtonian potential $\Phi$ diverges when calculated in a sphere of $r \rightarrow \infty$, making it necessary to impose a cutoff distance $r_\text{max}$. Chameleon-screened theories that motivate our investigation of $\Phi$ supply a natural choice for $r_\text{max}$, the Compton wavelength $\lambda_C$ of the scalar field that determines the effective range of the fifth force. Masses significantly further apart than $\lambda_C$ do not interact via the scalar coupling even if the field is otherwise unscreened. In $f(R)$ gravity, $\lambda_C$ is set by the background field value $f_{R0}$, which also determines the potential $\Phi_\text{crit}$ at which screening becomes operative~\citep{Hu_Sawicki,Cabre}:

\begin{equation}
\lambda_C \approx 32 \: \sqrt{f_{R0}/10^{-4}} \: \text{Mpc},
\end{equation}

\begin{equation}
\Phi_\text{crit}/c^2 \approx \frac{3}{2} f_{R0}.
\end{equation}

The $f_{R0}$ scale that may be probed by galaxies is therefore set by typical values of $\Phi$, which is a function of $r_\text{max}$. We use a fiducial value of $r_\text{max} = 10 \: \text{Mpc}$, corresponding to $f_{R0} \approx \Phi_\text{crit} \approx 10^{-5}$ for $\lambda_C = r_\text{max}$, but show results also for $r_\text{max} = 3 \: \text{Mpc}$. Since our other proxies ($\vec{a}$ and $K$) fall off more steeply with $r$ (and individual contributions to $\vec{a}$ sum vectorially), sources beyond $3 \: \text{Mpc}$ typically contribute little, making them fairly insensitive to the choice of $r_\text{max}$ as we demonstrate explicitly in Section~\ref{sec:distributions}. Their values are normally set by a few nearby objects.

\subsection{Primary source catalogue}
\label{sec:catalog}

We base our analysis on the 2M++ galaxy catalogue~\citep{2M++}, a synthesis of 2MASS, 6dF and SDSS data. This is an optimum catalogue for our purposes, for three reasons. First, it was designed with the goal of high all-sky completeness out to 200 Mpc. Complete sky coverage will prevent our having to restrict our maps to the footprint of a single survey, and 200 Mpc is around the largest distance that potential modified gravity signals are robustly measurable at present~\citep{Vikram}. Deeper surveys (e.g. the SDSS main sample) exist only over part of the sky. Second, it has a homogeneous limiting $K$-band magnitude ($m_K < 12.5$), which will facilitate our modelling of the contributions to the proxies from missing objects by means of an N-body simulation in Section~\ref{sec:sims}. Finally, the catalogue has already been used to estimate the smoothed density field in the survey volume~\citep{Lavaux}, which provides an important contribution to our proxies that is largely independent of those above.

To determine the values of $\Phi$, $\vec{a}$ and $K$ sourced by the 2M++ galaxies, we first estimate the mass distributions of their dark matter halos. We utilise the technique of ``halo abundance matching'' (AM), which maps galaxies to halos produced in an N-body simulation by postulating a nearly monotonic relationship between luminosity and some function of halo virial mass and concentration~\citep{Kravtsov,Conroy,Moster}. For a suitable scatter and function of halo properties, this has been shown to yield excellent agreement with galaxy clustering statistics (e.g.~\citealt{Reddick, Lehmann}) and moderate agreement with galaxies' internal dynamics (e.g.~\citealt{TG, DW15, DW16, D16}). We use the 2M++ $K$-band luminosity function measured in~\citet{2M++}, and \textsc{rockstar}~\citep{Rockstar_1} halos from the \textsc{darksky-400} simulation~\citep{DarkSky}, a $(400 \: \text{Mpc/h})^3$ box with $4096^3$ particles run with the \textsc{2hot} code~\citep{Warren13}. Our specific AM model will be that of~\citet{Lehmann}, which matches luminosity to $v_\text{vir}(v_\text{max}/v_\text{vir})^\alpha$, with best-fitting values $\alpha=0.6$ and uniform Gaussian scatter $\sigma_\text{AM} = 0.16$ dex. We caution that these parameters were derived using an $r$-band luminosity function from SDSS rather than a $K$-band one from 2M++, although we have verified that a basic counts-in-cells clustering statistic out to $10$ Mpc is consistent between the 2M++ dataset and our AM catalogue. This method enables us to generate an absolute $K$-band magnitude for each halo in the simulation box.\footnote{In principle there is further information in the group statistics of the 2M++ catalogue, for two reasons. First, using the group luminosity function rather than that of individual galaxies may improve the estimate of the halo masses derived from AM at the bright end. Second, the velocity dispersions of galaxies in a group provide complementary information to the luminosity on the distribution of dark matter mass. Folding this in would improve the precision of the results, although it is beyond the present scope of our work.}

Next, we next calculate the absolute magnitude $M$ of each galaxy in the 2M++ catalogue:

\begin{equation}
M = m - A_k(l,b) - k(z) + e(z) - D_L(z),
\end{equation}

\noindent where $m$ is the apparent magnitude, $A_k(l,b)$ describes dust absorption in the direction $(l,b)$, $k(z)$ is the k-correction due to redshifting of the spectrum, $e(z)$ is the correction for stellar population evolution, and $D_L$ is the luminosity distance. Following~\citet{2M++}, we take $k(z)=-2.1z$, $e(z)=0.8z$ and $A_k(l,b)=0.35E_\text{(B-V)}(l,b)$, and calculate the extinction factor $E_{(B-V)}$ from the dust map of~\citet{Schlegel}. We associate each 2M++ galaxy with the halo in the simulation closest to it in absolute magnitude. \red{We assume this halo to have a Navarro-Frenk-White (NFW) density profile~\citep{NFW},}

\begin{equation}
\rho(r) = \frac{M_\text{vir} c^2}{4 \pi r \, (R_\text{vir} + c r)^2 \, (\ln(1+c)-c/(1+c))},
\end{equation}

\noindent \red{with virial mass $M_\text{vir}$, virial radius $R_\text{vir}$ and concentration $c \equiv R_\text{vir}/r_\text{s}$ as output by \textsc{rockstar}.}

With this mass distribution in place, we are in position to calculate the proxy values sourced by each 2M++ object. For $\Phi$ and $\vec{a}$ we use standard forms for the NFW profile~\citep{Cole_Lacey}:

\begin{equation}
\Phi_\text{vis} = -\sum_\text{i} \: \frac{G M_\text{vir,i}}{r_\text{i}} \: \frac{\ln(1+\frac{c_\text{i} r_\text{i}}{R_\text{vir,i}})}{\ln(1+c_\text{i})-\frac{c_\text{i}}{1+c_\text{i}}},
\end{equation}

\begin{equation}
\vec{a}_\text{vis} = -\sum_\text{i} \: \frac{G M_\text{vir,i}}{r_\text{i}} \: \frac{\frac{1}{r_\text{i}} \ln(1+ \frac{c_\text{i} r_\text{i}}{R_\text{vir,i}}) - \frac{c_\text{i}/R_\text{vir,i}}{1+c_\text{i} r_\text{i}/R_\text{vir,i}}}{\ln(1+c_\text{i})-\frac{c_\text{i}}{1+c_\text{i}}} \hat{r_\text{i}},
\end{equation}

\noindent for source object $i$ at distance $r_\text{i}$ from the test point, where $\hat{r}$ points from the source halo centre to the test point. The subscript `vis' denotes that these contributions to the proxies derive from objects visible to the 2M++ survey; further contributions will be described in Sections~\ref{sec:sims} and~\ref{sec:lavaux}. We measure curvature using the Kretschmann scalar $K \equiv (R^{\alpha \beta \gamma \delta} R_{\alpha \beta \gamma \delta})^{1/2}$, both because it is non-zero in vacuum and to promote compatibility with~\citet{Baker}, who find it to be of use in synthesising laboratory, astrophysical and cosmological constraints on gravity. While in detail the Kretschmann scalar for a set of point masses is nonlinear, we find the error incurred by approximating it as the linear sum of individual contributions to be small. We assume also that $K$ can be adequately calculated by treating each source halo as a point mass at its centre. As we make no attempt to estimate the accuracy of this approximation, $K$ ought not to be considered more than an order-of-magnitude estimator of the true curvature.

\begin{equation}
K_\text{vis} = \sum_\text{i} \: \sqrt{48} \: \frac{G M_\text{vir,i}}{r^3}.
\end{equation}

\subsection{Restoring missing halos}
\label{sec:sims}

The 2M++ catalogue clearly does not include all mass within 200 Mpc, and hence the calculation of Section~\ref{sec:catalog} underestimates $\Phi$, $\vec{a}$ and $K$. Part of the missing mass resides in halos hosting galaxies too faint to be observed in the 2M++ survey yet sufficiently massive to be well-resolved in the N-body simulation to which we have already had recourse. This mass can be filled in by calibrating with that simulation as follows:

\begin{enumerate}

\item{} \red{Define a well-resolved halo as one containing at least $1000$ particles within $R_\text{vir}$, which for the \textsc{darksky-400} simulation equates to a mass cut $M_\text{vir} > 7.63 \times 10^{10} \: h^{-1} \: M_\odot$. Above this threshold halos have reliable concentration measurements~\citep{Diemer_Kravtsov}, allowing Eqs. 5-7 to be evaluated accurately, and the halo mass function is complete~\citep{Reed}, ensuring that all such halos are identified by the halo finder.}

\item{} Using the absolute magnitude assigned to each halo from AM, and assuming the observer to be at the centre of the box, determine the apparent magnitude from Eq. 3.\footnote{Although the N-body box is not intended to represent the actual local universe, we nevertheless use the angular dependence of the~\citet{Schlegel} dust map to calculate the extinction at the position of each halo. As this term is small, a different angular dependence would have little effect.} Thus split the halos into those that would be visible in the 2M++ catalogue ($m<12.5$) and those that would not ($m>12.5$).

\item{} At the position of each well-resolved halo, calculate $\Phi$ from all other well-resolved halos within 10 Mpc ($\Phi_\text{hal}$), and from the visible ones alone ($\Phi_\text{vis}$). (It is advisable to use the positions of the halos themselves rather than random points in the box because the latter would not reproduce the clustering properties of real galaxies. In particular, low density regions would be much more prominently represented.) Take $c_\Phi \equiv \Phi_\text{hal}/\Phi_\text{vis}$ as the multiplicative ``correction factor'' required to map the latter potential onto the former.\footnote{Note that due to redshift incompleteness in the 2M++ catalogue, as discussed in~\citet{2M++}, fewer objects are actually observed than pass the magnitude cut, an effect more pronounced in some directions than others. This causes our correction factors to be underestimated, as we assume that all objects with $m<12.5$ contribute to the observable proxies. However, as the survey achieved high completeness in all regions above the galactic disk~\citep{2M++}, this effect is insignificant there relative to the other sources of uncertainty in our analysis.} \red{Repeat to obtain the corresponding correction factors for acceleration and curvature, $c_a \equiv |\vec{a}_\text{hal}|/|\vec{a}_\text{vis}|$ (describing the change in the magnitude of $\vec{a}$), $c_\theta \equiv \frac{\vec{a}_\text{hal} \cdot \vec{a}_\text{vis}}{|\vec{a}_\text{hal}| |\vec{a}_\text{vis}|}$ (describing the change in the direction of $\vec{a}$) and $c_K \equiv K_\text{hal}/K_\text{vis}$. Denote the set of correction factors by $\vec{c} \equiv \{c_\Phi, c_a, c_\theta, c_K\}$. Record also the distance to the halo in question ($d$), the number of visible objects within 10 Mpc of it ($N_{10}$), and $\Phi_\text{vis}$. We find these observables to correlate strongly with $\vec{c}$, and hence will use them to predict it.}

\item{} \red{Partition the \{$c_\Phi$, $c_a$, $c_\theta$, $c_K$, $d$, $N_{10}$, $\Phi_\text{vis}$\} space into discrete cells and populate it with the halos in the box. This estimates their joint frequency distribution. We sample logarithmically in $c_K$ and $\Phi_\text{vis}$, take $35$ bins in each direction, and choose the upper and lower limits to enclose at least $95$ per cent of the simulation points.}

\item{} \red{For each test galaxy in the real universe, estimate $\vec{c}$ by drawing random values for each of its components from this distribution at the \{$d$, $N_{10}$, $\Phi_\text{vis}$\} of that galaxy. This measures $\vec{c}$ conditional on \{$d$, $N_{10}$, $\Phi_\text{vis}$\}, and retains the full covariance among its components.}

\item{} Estimate the total potential and curvature values for the test galaxy by multiplying the estimates from the visible mass alone by the corresponding correction factors, i.e. $\Phi_\text{hal} = c_\phi \times \Phi_\text{vis}$ and $K_\text{hal} = c_K \times K_\text{vis}$. For acceleration, first multiply $|\vec{a}_\text{vis}|$ by $c_a$, then randomly rotate the resulting vector through $c_\theta$.

\end{enumerate}

\subsection{Modelling the remaining missing mass}
\label{sec:lavaux}

\red{The combination of halos hosting visible galaxies and those well-resolved in our simulation accounts for only $\sim 10$ per cent of the universe's mass; the remainder is located either in smaller halos or outside any virialised structure (e.g.~\citealt{vanDaalen}). The analysis of Section~\ref{sec:sims} therefore continues to underestimate the proxies. To model the remaining mass we use the results of~\citet{Lavaux}, who apply 2LPT in a Bayesian framework to derive the probability distribution for the $z=0$ density field consistent with the number density and peculiar velocities of the 2M++ galaxies. This is done on a 3D grid of spacing $2.3 \: h^{-1} \: \text{Mpc}$ and therefore provides an accurate estimate of the power spectrum on scales $k \lesssim 1 \: h \: \text{Mpc}^{-1}$.}

\red{Although this method reconstructs approximately $100$ per cent of the mass in the 2M++ volume, the smoothing scale of $\sim 3.4$ Mpc means that the screening proxies cannot be estimated from this mass alone; their scaling with $r^{-n}$ makes them sensitive to smaller-scale overdensities such as we have already modelled in Sections~\ref{sec:catalog} and~\ref{sec:sims}. We estimate the proxy values due to \emph{all} mass by summing the contributions from the resolved halos and 2M++ galaxies, derived in Section~\ref{sec:sims}, to those from the mass field derived by smoothly interpolating the density grid of~\citet{Lavaux} at a representative point in their MCMC chain.\footnote{The results are not significantly altered by choosing a different high-likelihood point from the chain, although see Section~\ref{sec:error} for further discussion of this source of uncertainty.} As this method double-counts a small fraction of the mass, it systematically overestimates $\Phi$, $a$ and $K$;\red{\footnote{\red{This provides further motivation for limiting the contribution in Section~\ref{sec:sims} to halos with $>1000$ particles: including smaller halos (or using the particles themselves) would double-count mass more egregiously.}}} our results for the minimum degree of screening -- and hence the strongest limits on scalar-tensor theory screening thresholds afforded in principle by galaxy observations -- are therefore conservative. The two contributions do however provide power on largely complementary scales: the halos may be seen as filling in the power from the 1-halo term that is washed out by the $3.4$ Mpc smoothing. We provide further discussion of this difficulty, including means by which it may be overcome, in Section~\ref{sec:error}.}

\vspace{5 mm}

\noindent Our code for performing all the operations of this section is publicly available at \url{https://www2.physics.ox.ac.uk/contacts/people/desmond}.

\section{Results}
\label{sec:results}

In a real application, our method would be used to evaluate $\Phi$, $\vec{a}$ and $K$ at the positions of galaxies with measured values for potential modified gravity signals. Since we do not yet possess such a sample, we instead show results at the positions of galaxies in our source catalogue, which at least sample realistic galaxy environments. In this section we will investigate the impact of unseen halo mass and the reliability of our method of estimating it (Section~\ref{sec:unseen}), the relative contributions of the three mass components (Section~\ref{sec:relative}), and the overall proxy distributions (Section~\ref{sec:distributions}).

\subsection{Modelling unseen mass}
\label{sec:unseen}

\begin{figure*}
  \subfigure[]
  {
    \includegraphics[width=0.48\textwidth]{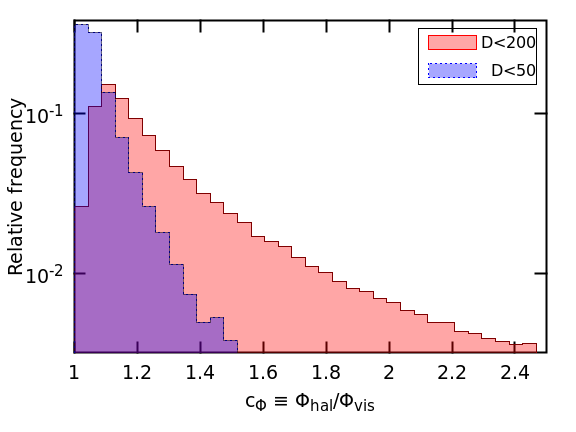}
    \label{fig:CorrFac_V}
  }
  \subfigure[]
  {
    \includegraphics[width=0.48\textwidth]{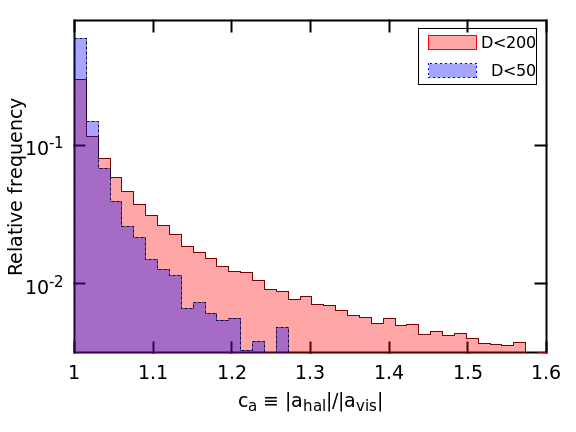}
    \label{fig:CorrFac_a}
  }
  \subfigure[]
  {
    \includegraphics[width=0.48\textwidth]{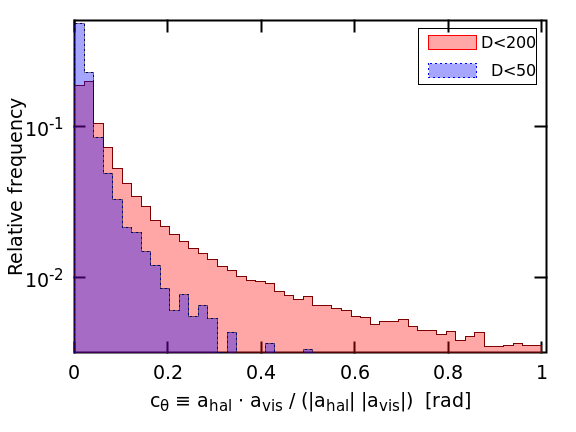}
    \label{fig:CorrFac_ang}
  }
  \subfigure[]
  {
    \includegraphics[width=0.48\textwidth]{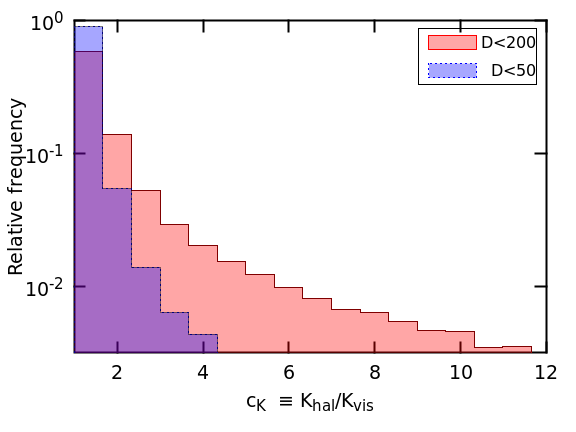}
    \label{fig:CorrFac_c}
  }
  \caption{The distributions of the correction factors relating the contributions to $\Phi$ ($c_\Phi$, Fig.~\ref{fig:CorrFac_V}), $\vec{a}$ ($c_a$, Fig.~\ref{fig:CorrFac_a} and $c_\theta$, Fig.~\ref{fig:CorrFac_ang}) and $K$ ($c_K$, Fig.~\ref{fig:CorrFac_c}) from 2M++ objects (subscript `vis') to those from all halos (`hal'). $c_a$ describes the change in magnitude of the acceleration vector, and $c_\theta$ the angle through which it is rotated (see Section~\ref{sec:sims}). The correction factors typically increase with distance from the observer.}
  \label{fig:CorrFacs}
\end{figure*}

We begin with our determination of the correction factors $c_\Phi$, $c_a$, $c_\theta$ and $c_K$: histograms of these quantities are shown in Figure~\ref{fig:CorrFacs} for galaxies within 200 and 50 Mpc. The averages and standard deviations of these distributions are reported in the second and third columns of Table~\ref{tab:table}, while the fourth contains the average dispersions at fixed values of $d$, $N_{10}$ and $\Phi_\text{vis}$. These measure the uncertainties on our estimate of $\vec{c}$ using the method of Section~\ref{sec:sims}. The most significant features of these results are the following:

\begin{table*}
  \begin{center}
    \begin{tabular}{l|c|c|c|c|}
      \hline
      \textbf{Correction factor} & \textbf{Median} & \textbf{Mean} & \textbf{St. dev.} & \textbf{St. dev. after calibration}\\ 
      \hline
\rule{0pt}{3ex}
      $c_\Phi$ 		           & \red{1.1}		& \red{1.5}        & \red{1.4}	   & \red{0.18}\\
\rule{0pt}{3ex}
      $c_a$		           & \red{1.1} 	        & \red{3.9}        & \red{58}        & \red{0.53}\\
\rule{0pt}{3ex}
      $c_\theta$                   & \red{0.07} 	& \red{0.29}       & \red{0.51}        & \red{0.33}\\
\rule{0pt}{3ex}
      $\log_{10}(c_K)$	           & \red{0.09} 	& \red{0.45}       & \red{0.83}         & \red{0.14}\\
      \hline
     \end{tabular}
  \caption{Statistics describing the distributions of correction factors relating proxy values from 2M++ objects to those from all halo mass. ``Calibration'' refers to our procedure for estimating the correction factors from \{$d$, $N_{10}$, $\Phi_\text{vis}$\} (Section~\ref{sec:sims}), which can be seen to reduce their uncertainties. Note that the first three columns include all well-resolved halos in the simulation, while the last pertains to the correction factors actually assigned at the positions of the 2M++ galaxies (as in Fig.~\ref{fig:CorrFacs}).}
  \label{tab:table}
  \end{center}
\end{table*}

\begin{enumerate}

\item{} As the correction factors are not negligible, they are important in any calculation of $\Phi$, $\vec{a}$ or $K$. Tails towards high values indicate environments dominated by unseen mass.

\item{} On average, $c_K > c_a > c_\Phi$, and the variances of the distributions increase also in that order (this is due to contributions below the relative frequency cutoff in Fig.~\ref{fig:CorrFacs} and hence not visible there). This is the order of decreasing sensitivity of the proxy to the separation $r$ of source and test point: $|\vec{a}|$ and $K$ are more dependent than $\Phi$ on nearby low-mass objects which are less likely to be included in 2M++. This makes our method most reliable for determining $\Phi$ and least for $K$, as the latter is more likely to be dominated by a single faint object. Indeed, at some test points $K_\text{hal}$ exceeds $K_\text{vis}$ by more than an order of magnitude (note the logarithm in the final row of Table~\ref{tab:table}), making a method based on a magnitude-limited source galaxy sample unreliable.

\item{} The widths of the $\vec{c}$ distributions are significantly smaller after calibrating with \{$d$, $N_{10}$, $\Phi_\text{vis}$\} than before, showing that this method improves the precision of our proxy calculations. The remaining variance contributes to the final uncertainty in our maps, as we discuss further in Section~\ref{sec:error}.

\item{} The correction factors are larger further away since a smaller fraction of objects passes the 2M++ magnitude cut.

\end{enumerate}

\subsection{Relative contributions to the proxies}
\label{sec:relative}

\begin{figure*}
  \subfigure[]
  {
    \includegraphics[width=0.318\textwidth]{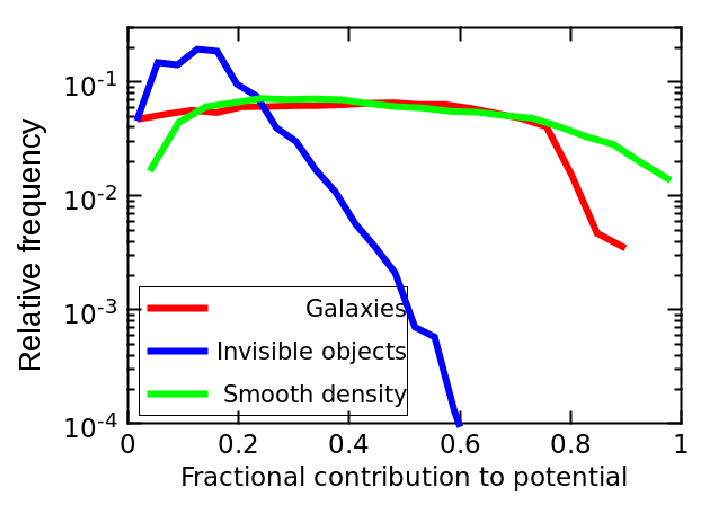}
    \label{fig:V_comp}
  }
  \subfigure[]
  {
    \includegraphics[width=0.3\textwidth]{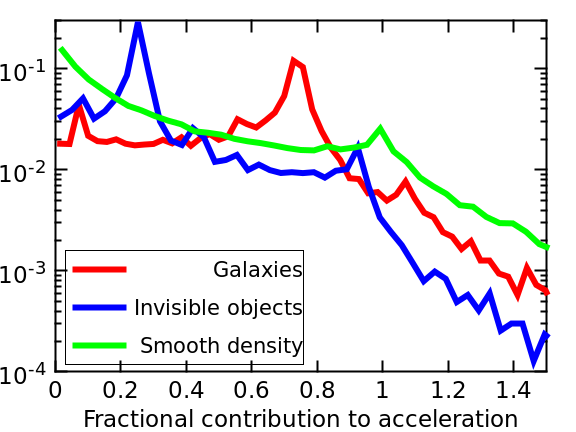}
    \label{fig:a_comp}
  }
  \subfigure[]
  {
    \includegraphics[width=0.3\textwidth]{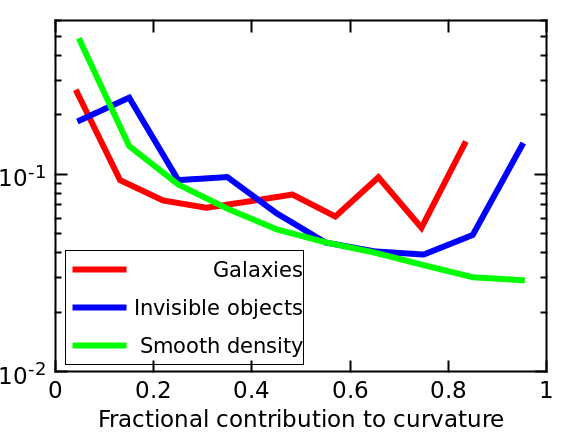}
    \label{fig:c_comp}
  }
  \caption{The relative contributions to $\Phi$, $|\vec{a}|$ and $K$ of the 2M++ galaxies, unseen halo mass and smoothed density field. Since contributions to $\vec{a}$ sum vectorially, the magnitude of the acceleration from a given component may be larger than that of the total.}
  \label{fig:Comparison}
\end{figure*}

Next we show in Figure~\ref{fig:Comparison} the relative contributions to the proxies from the observed 2M++ galaxies, the halos from the N-body box and the smoothed density field. That they are roughly the same order of magnitude shows that none are negligible. In the case of $\Phi$, the fractional contribution of the 2M++ galaxies is roughly uniform up to 80 per cent, indicating that in few cases is a fraction of $\Phi$ greater than this sourced by those galaxies alone. The invisible halos in the simulation typically contribute $\sim 20\%$, and rarely more than $50\%$. Some test galaxies have no or very halos or 2M++ objects within 10 Mpc, making both $\Phi_\text{vis}$ and $\Phi_\text{hal}$ very small; the potential for these objects effectively derives purely from the smoothed density field.

Interpretation of the relative contributions of the three components to $\vec{a}$ (Fig.~\ref{fig:a_comp}) is complicated by the fact that they  do not sum directly at a given test point, but rather vectorially. Depending on the phases, this means that the magnitudes of the acceleration due to any one of the components may be larger than the total acceleration, making the fractional contribution greater than 1. The 2M++ part peaks at around $75\%$, and most often the remaining $25\%$ (with roughly aligned directions) comes from unseen halos. Since mass in the low-frequency modes of the smoothed density field is by definition fairly homogeneous (and that in the zero-frequency mode entirely so), it tends to have little effect on the acceleration as contributions from opposite sides of the test galaxy roughly cancel out. That said, the small fraction of test points with little 2M++ or unseen halo mass within 10 Mpc have accelerations dominated by the low-frequency modes, as evidenced by the peak in the green curve at $\sim1$.

We find the curvature typically to be dominated by a single component, most often the 2M++ galaxies (Fig.~\ref{fig:c_comp}). Since $K$ falls more steeply with $r$ than $\Phi$ or $\vec{a}$, and hence samples on average a smaller volume around a given test point, it is less affected by mass in the smoothed density field.

\subsection{Distributions of gravitational variables}
\label{sec:distributions}

\begin{figure*}
  \subfigure[]
  {
    \includegraphics[width=0.3\textwidth]{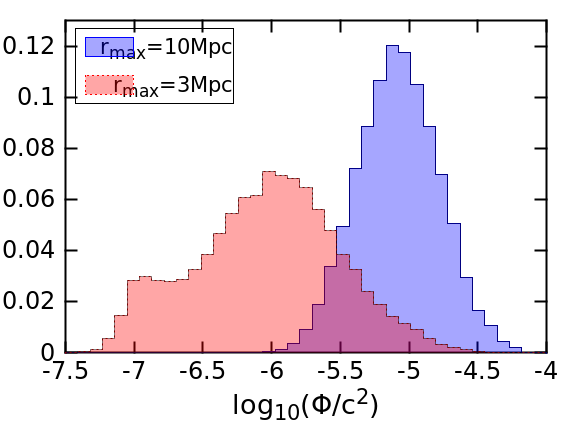}
    \label{fig:V}
  }
  \subfigure[]
  {
    \includegraphics[width=0.3\textwidth]{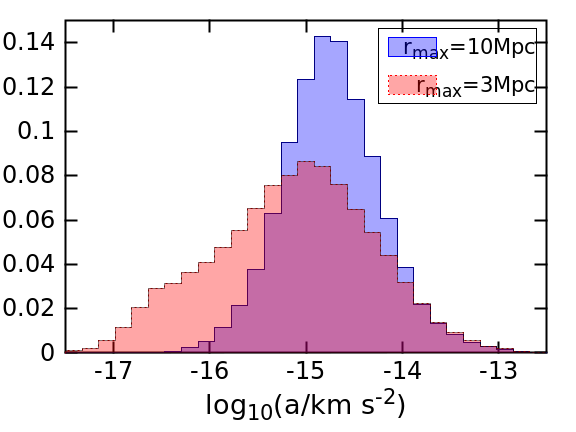}
    \label{fig:a}
  }
  \subfigure[]
  {
    \includegraphics[width=0.3\textwidth]{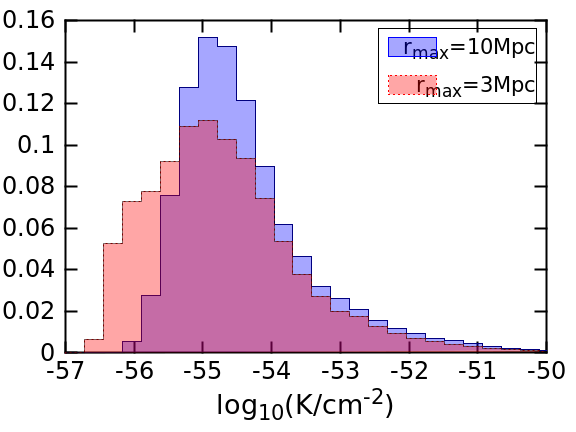}
    \label{fig:c}
  }
  \caption{Distributions of $\Phi$, $|\vec{a}|$ and $K$ at the locations of the 2M++ galaxies, calculated within a sphere of $10 \: \text{Mpc}$ (blue) or $3 \: \text{Mpc}$ (red) around each test point. The ranges of these variables set the potential of galaxies to extend the scope of probes of modified gravity.}
  \label{fig:Proxies}
\end{figure*}

Figure~\ref{fig:Proxies} shows the distributions of $\Phi$, $\vec{a}$ and $K$ across our test objects, including source objects up to $r_\text{max} = 10 \: \text{Mpc}$ (blue) or 3 Mpc (red). As mentioned in Section~\ref{sec:intro}, the choice of $r_\text{max}$ is more important for $\Phi$ than $a$ or $K$ due to its weaker fall off with the separation of test and source object: most of the mass contributing to $a$ and $K$ is located within 3 Mpc. We find the proxy distributions not to depend significantly on the masses of the test galaxies, and thus they provide an indication of the range of values probed by any galaxy sample without explicit selection on environment.

\begin{figure*}
  \subfigure[$\Phi$]
  {
    \includegraphics[width=0.3\textwidth]{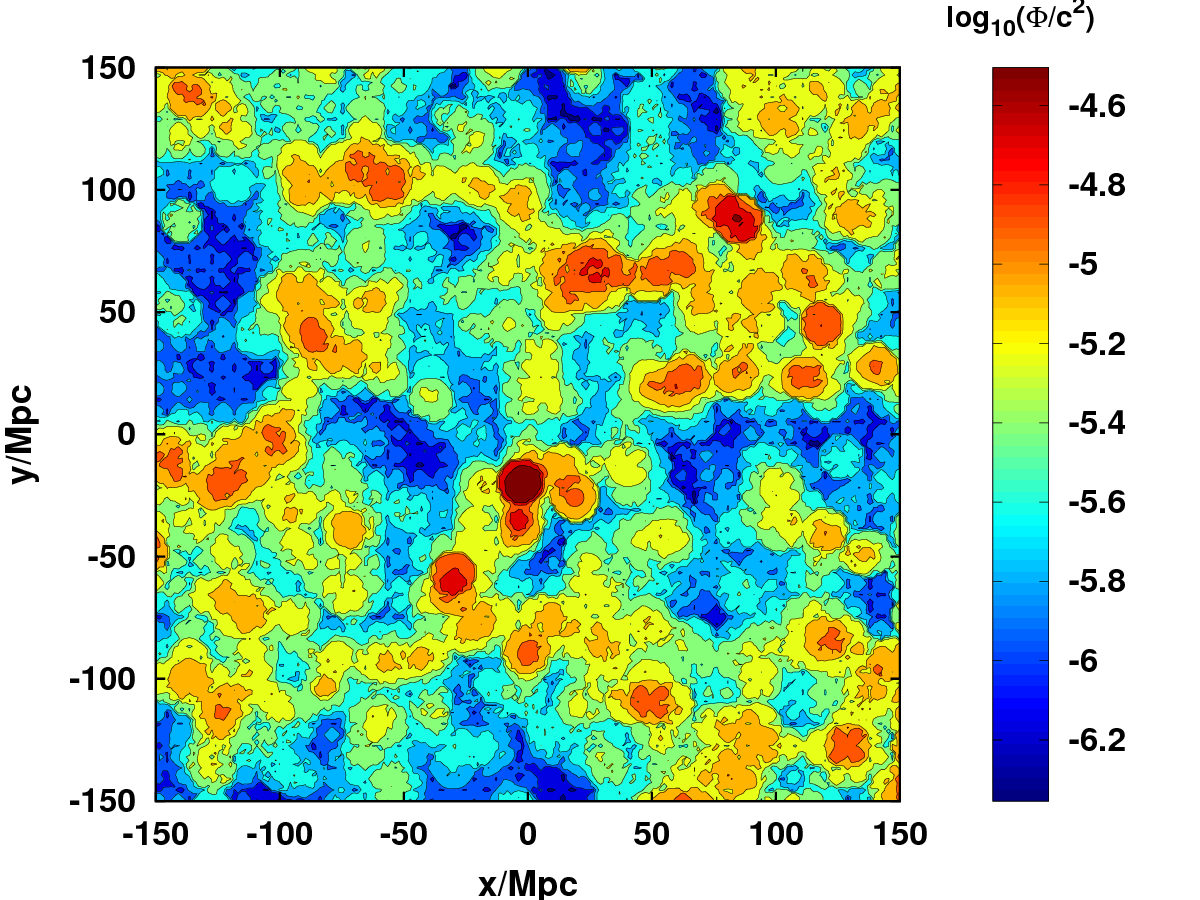}
    \label{fig:map_V}
  }
  \subfigure[$|\vec{a}|$]
  {
    \includegraphics[width=0.3\textwidth]{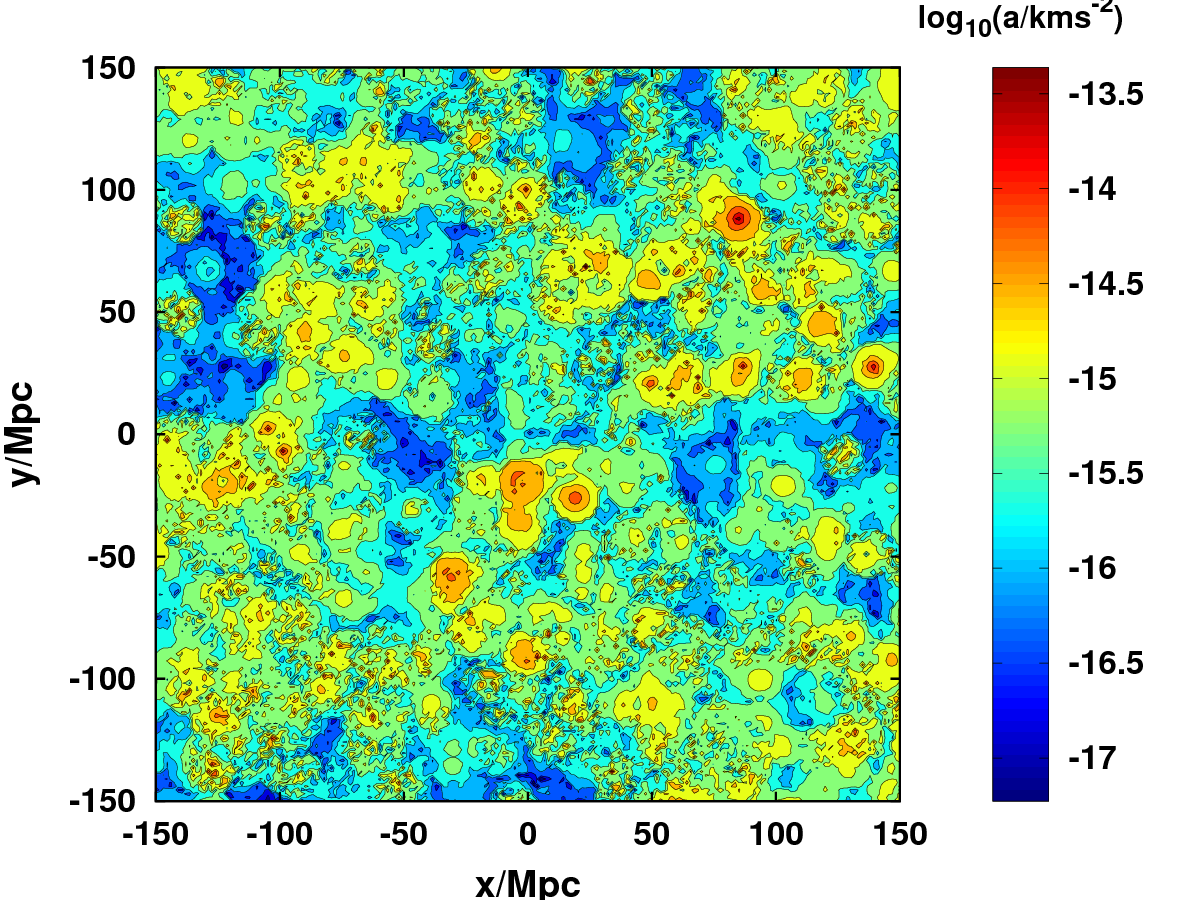}
    \label{fig:map_a}
  }
  \subfigure[$K$]
  {
    \includegraphics[width=0.3\textwidth]{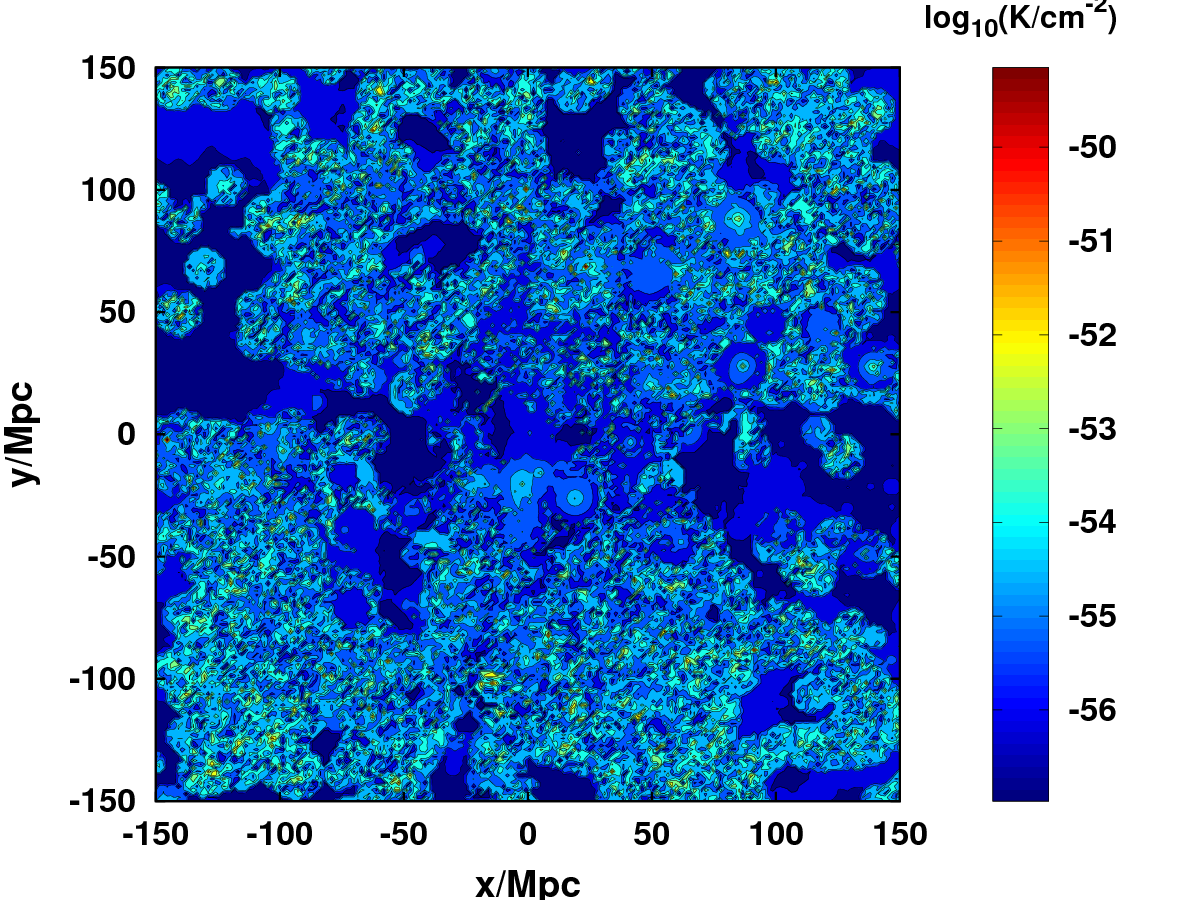}
    \label{fig:map_c}
  }
  \caption{Contour plots of $\Phi$, $|\vec{a}|$ and $K$ across a $300 \: \text{Mpc} \times 300 \: \text{Mpc}$ slice of the local universe. The Milky Way is located at $x=y=0$.}
  \label{fig:Maps}
\end{figure*}

Finally, in Figure~\ref{fig:Maps} we display the proxy values across a $300 \: \text{Mpc} \times 300 \: \text{Mpc}$ slice of the local universe, again using $r_\text{max}=10$ Mpc. The Milky Way is located at $x=y=0$, and the sampling resolution is 1.5 Mpc. The increasingly detailed level of structure in these maps when moving from $\Phi$ to $a$ to $K$ reflects again the greater sensitivity of the proxy to the separation $r$ of source and test points. Their approximate uniformity over the entire area attests to the accuracy of our method for restoring missing halos, of which there are more further away.

These maps reveal promising regions in which to search for modified gravity; for example, regions of deep blue in Fig.~\ref{fig:map_V} have very low $\Phi/c^2$ (down to few $\times10^{-7}$), allowing any galaxies within them to remain unscreened in chameleon theories with small background scalar field values. Conversely, regions of deep red in Fig.~\ref{fig:map_a} have particularly high accelerations, making them most likely to harbour external field-dominated galaxies in MOND.

\section{Discussion}
\label{sec:discussion}

In this section we discuss the implications of our results for galaxy-scale tests of modified gravity (Section~\ref{sec:modgrav}), and then describe the major uncertainties in our maps and the potential for future improvement (Section~\ref{sec:error}).

\subsection{Applications to modified gravity}
\label{sec:modgrav}

We begin with a few general inferences about the usefulness of galaxy-scale tests of gravity from the distributions of our proxy values (Fig.~\ref{fig:Proxies}).

\begin{enumerate}

\item{} In chameleon scalar-tensor theories, the value of $\Phi$ marking the transition between the screened and unscreened regimes is set by the background value of the scalar field~\citep{Hu_Sawicki}. A test of screening involves a comparison between putatively screened and unscreened samples, and hence the lowest value of $\Phi$ determines the strength of the constraint that may in principle be placed on the theory. Figure~\ref{fig:V} suggests that a significant number of galaxies are likely to be environmentally unscreened for $\Phi_\text{crit}/c^2 = 10^{-7}$ and $r_\text{max} = \lambda_C = 1 \: \text{Mpc}$, giving galaxy-scale tests the potential to probe $f_{R0}$ to at least the $10^{-7}$ level (see Eqs. 1 and 2). Although our results differ in detail from those of~\citet{Cabre}, due to our more sophisticated treatment of missing mass, we share this optimistic conclusion.

\item{} Typical external accelerations are approximately a few $\times 10^{-15} \: \text{km} \, \text{s}^{-2}$, with a range of $\sim10^{-16} - {\rm few} \times 10^{-14}$ (Fig.~\ref{fig:a}). Given the relatively small difference between the red and blue curves in that figure, especially at the high end, this is not likely to change much if $r_\text{max}$ was extended. In MOND-type models, a galaxy enters the external field-dominated regime when the external acceleration $a_\text{ex}$, due to surrounding mass, exceeds the acceleration $a_\text{in}$ generated by the galaxy itself, provided $a_\text{in}$ is less than the characteristic acceleration $a_0\approx 1.2 \times 10^{-13} \: \text{km} \, \text{s}^{-2}$. Thus our distribution of $a_\text{ex}$ indicates the maximum internal accelerations -- and hence surface densities -- that galaxies can have for the external field to be important. A typical value $a_\text{ex} \sim {\rm few} \times 10^{-15} \: \text{km} \, \text{s}^{-2}$ would correspond to surface density $\Sigma \approx 20 \: M_\odot \: \text{pc}^{-2}$, which is very small even for a low surface brightness galaxy. Thus only unusually high $a_\text{ex}$ values would appreciably impact dynamics. Galaxies with $a_\text{ex} > a_0$ (of which Fig.~\ref{fig:a} suggests there are very few) should exhibit fully Newtonian behaviour. We caution however that we have measured the proxies only at the positions of the 2M++ galaxies; more promising galaxies for tests of the external field effect are dwarfs very close to a single massive object which provides the bulk of the external acceleration (such as the dSphs of the Milky Way and M31;~\citealt{McGaugh_EFE, McGaugh_EFE_2}). It is unlikely that these would be included among our test galaxies due to high apparent magnitudes at any distance beyond the local group. At the positions of such galaxies located by other means, however, our map would be expected to report $a_\text{ex}$ accurately provided the galaxies sourcing the field were included in the 2M++ catalogue.

At the other end of the scale, by spherical symmetry (and unlike for $\Phi$ and $K$), the smoothed density field does not provide an irreducible background value of $\vec{a}$, and hence there is no fundamental reason why regions of arbitrarily low $\vec{a}$ could not exist. This enables screening mechanisms characterised by acceleration (e.g. kinetic) to be tested in principle to arbitrary precision in galaxies, provided a sample in sufficiently remote environments could be compiled.

\item{} The external curvature is typically around $K = \text{few} \: \times 10^{-54} \: \text{cm}^{-2}$, with a range of roughly $10^{-56} - 10^{-51}$. This sets the precision with which Vainshtein screening may be tested at the galaxy scale. These values are at the low end of the ``curvature desert'' identified in~\citet{Baker}, enabling galaxies potentially to inhabit any part of this region when their internal curvature contributions are added.\footnote{\citet{Baker} consider only the internal contributions. For models in which the important variable is the total proxy value, our results should therefore be considered summed to the galaxy lines in their fig. 1.} As $K$ falls off rapidly with $r$, these contributions will typically greatly exceed the environmental ones considered here. Note also that the lower limit of our distribution, $K \approx 10^{-56} \: \text{cm}^{-2}$, coincides roughly with the ``Lambda'' line of~\citet{Baker}'s fig. 1, which is approximately the curvature of the cosmological background.

\end{enumerate}

More detailed or model-specific conclusions concerning modified gravity will require a test galaxy sample for which measurements of potential modified gravity signals have independently been made. Correlations of these signals with the proxies determined from our maps (plus the internal contributions from the galaxies themselves) may then be used to search for a transition to a new gravitational regime or test specific theories for which these proxies are important.

Such signals will depend on the theory in question. In scalar-tensor theories, the stars in unscreened galaxies may self-screen, leading to differences between their kinematics and those of the gas and dark matter. In particular, the stars would fall more slowly in an external field, leading to displacements between the centroids of stellar and H\textsc{i} disks preferentially aligned with the external acceleration, and to an enhancement of H\textsc{i} over H$\alpha$ rotation curves. The offset between halo and disk centres may then cause stellar disks to warp, and both the photometry and kinematics of the disk to develop asymmetries~\citep{Jain_Vanderplas,Vikram}.

In MOND, the value of $|\vec{a}|$ sets the ratio of a system's dynamical to baryonic mass, and also the shape of the rotation or velocity dispersion profile. In galaxies with internal acceleration $a_\text{in} \ll a_0$, rotation curves would be flat for $a_\text{ex} < a_\text{in}$ (the standard deep-MOND regime), but Keplerian for $a_\text{ex} > a_\text{in}$ (the external field-dominated regime;~\citealt{Famaey_McGaugh}). The galaxy's total mass discrepancy, measured by $\mathcal{D} \equiv \frac{R V^2}{M_\text{bar} G}$, would then be $\sim a_0/a_\text{ex}$ for $a_\text{ex} < a_0$ (the quasi-Newtonian regime) and $1$ for $a_\text{ex} > a_0$ (the Newtonian limit). The correlation of $\mathcal{D}$ with $a_\text{ex}$ -- the external-field analogue of the mass discrepancy--acceleration relation~\citep{MG99, Lelli} -- would have a characteristic shape and a feature at $a_\text{ex} = a_0$. 

Finally, we note that our maps may also find application in the study of galaxy formation. While correlations between $\Phi$, $a$ or $K$ and galaxy signals may be evidence for novel gravitational physics, they would more likely attest to galaxy growth or evolution effects in $\Lambda$CDM. They may therefore be useful in studies of quenching, conformity, satellite--host interactions and the environment dependence of the galaxy--halo connection. These phenomena will also constitute important systematics in tests of gravity.

\subsection{Error analysis and future prospects}
\label{sec:error}

\subsubsection{Sources of uncertainty}

Here we describe the most significant uncertainties in our maps, \red{in roughly decreasing order of importance}. This will give an indication of the areas on which future analyses should focus to best improve precision, and the ways in which upcoming galaxy surveys will be able to contribute. \red{The parameters that form the input to our analysis, and the distributions from which they are drawn, are summarised in Table~\ref{tab:params}.}

\begin{enumerate}

\item{} We estimated the correction factors required to convert the proxies sourced by visible mass to values due to all resolved halos on a point-by-point basis using their correlations with distance, the number of 2M++ objects within 10 Mpc, and the potential from these objects. However, as shown in Table~\ref{tab:table}, significant uncertainties remain, especially for $|\vec{a}|$ ($\sim200\%$) and $K$ ($\sim0.7$ dex). These uncertainties increase with distance. Since this uncertainty derives from the fraction of objects included in the galaxy survey, it is best reduced by employing a deeper survey. Over a third of the sky the magnitude limit may already be improved by using the SDSS main sample, with a Petrosian $r$-band magnitude limit of $17.7$, and the Taipan survey will soon provide similar data in the South. Within the next decade, the Dark Energy Spectroscopic Instrument (DESI) will push around two magnitudes deeper, directly locating an even greater proportion of the local universe's mass. \red{Indeed, we find that for a DESI-like catalogue with a limiting $r$-band magnitude of $19.7$ and average $r-K$ colour of $2.7$ (e.g.~\citealt{Bell}), essentially all halos with $>1000$ particles in our simulation would be expected to host visible galaxies out to 200 Mpc, all but eliminating the need for correction factors.} We refer the reader to~\citet{Jain_Surveys},~\citet{Jain_Khoury} and~\citet{Jain_NovelProbes} for further discussion of survey requirements and optimisation for the purposes of gravitational physics.

\item{} The analysis of~\citet{Lavaux} derives a probability distribution for the smoothed density field of the local universe from the positions and velocities of the 2M++ galaxies. The uncertainty that the width of this distribution propagates into our maps may be estimated by averaging over the results from all field configurations in the posterior, effectively marginalising over the possibilities. Although we leave this to future work, we have verified by direct substitution of several high-likelihood configurations that at typical locations the resulting uncertainties in $\Phi$, $a$ and $K$ are subdominant to those described above. We caution however that this uncertainty is most important in void regions containing few massive halos, which are the best places to search for screening.

\item{} Galaxies of given luminosity may reside in halos with a range of properties, so that the observed distribution of light does not uniquely determine the distribution of total mass. Thus the 2M++ galaxy sample corresponds to a range of possible $\Phi_\text{vis}$, $\vec{a}_\text{vis}$ and $K_\text{vis}$ fields, of which we show simply example realisations. \red{Within the AM framework, this derives from the combination of a statistical uncertainty due to the range of possible halo $M_\text{vir}$ and $c$ for a galaxy of given luminosity, quantified by $\sigma_\text{AM}$, and a potential systematic error due to uncertainties in the model parameters $\alpha$ and $\sigma_\text{AM}$ themselves. Approximately, $\sigma_\text{AM} = 0.16$~\citep{Lehmann} produces a $\sim0.16$ dex uncertainty in the mass of the halo of a galaxy of given luminosity, and hence in the values of $\Phi$, $a$ and $K$ sourced by that halo. This uncertainty is most significant for test points surrounded by many 2M++ galaxies.} In addition, baryonic effects may cause halo profiles to differ from the NFW form that we assume (e.g.~\citealt{Blumenthal, Governato, Pontzen}).

\item{} Uncertainties in the measured magnitude, position and distance of each source object propagate into uncertainties in our maps.

\end{enumerate}

\red{These uncertainties may all be accounted for in tandem by generating many Monte Carlo realisations of our maps with each input parameter drawn randomly from its set of allowed values, thus producing a posterior distribution of $\Phi$, $a$ and $K$ at each test point. These may be further propagated into constraints on modified gravity parameters given a full inference framework and a test galaxy sample. This will be the subject of future work.}

\begin{table}
  \begin{center}
    \begin{tabular}{l|l|}
      \hline
      \textbf{Input} & \textbf{Origin of parent distribution}\\
      \hline
\rule{0pt}{3ex}
      $\vec{c} \equiv \{c_\Phi, c_a, c_\theta, c_K\}$ 		& Regions of N-body simulations with \\&given distributions of visible mass\\
\rule{0pt}{3ex}
      Smoothed density field	           & Galaxy number density and peculiar\\& velocity fields~\citep{Lavaux}\\
\rule{0pt}{3ex}
      $P(M_\text{vir}, c|L; \alpha, \sigma_\text{AM})$          & Abundance matching mock with \\&fixed halo proxy and scatter\\
\rule{0pt}{3ex}
      $P(\alpha, \sigma_\text{AM})$	   & Galaxy clustering analysed by AM \\&\citep{Lehmann}\\
\rule{0pt}{3ex}
      $P(L_\text{t}, \vec{x}_\text{t}|L_\text{o}, \vec{x}_\text{o})$	   & Galaxy survey measurements\\
      \hline
     \end{tabular}
  \caption{\red{Parameters forming the input to our calculations of $\Phi$, $\vec{a}$ and $K$, and the source of the distributions from which they are drawn. In this work we take approximate maximum-likelihood values for each; in future work full uncertainties will be estimated by Monte Carlo marginalisation. Subscripts `t' and `o' in the final row denote true and observed values respectively.}}
  \label{tab:params}
  \end{center}
\end{table}

\subsubsection{Overestimation of total mass}

\red{As discussed in Section~\ref{sec:lavaux}, including halos in addition to the density field of~\citet{Lavaux} overestimates the total mass in the universe by $\sim10$ per cent, and hence $\Phi$, $a$ and $K$ by a comparable factor. Here we describe two methods by which this discrepancy may be eliminated.}

\begin{enumerate}

\item{} \red{First, estimate for each test point the total mass within the surrounding volume of interest (in our fiducial case a sphere of radius $10$ Mpc) using a smoothed field akin to~\citet{Lavaux}. This is reasonable provided the smoothing scale is less than the size of the volume of interest. Then estimate the true configuration of that mass as follows. A part is associated with visible galaxies (e.g. in 2M++), and may therefore be situated more precisely in halos around those galaxies specified by inverse AM, as in Section~\ref{sec:catalog}. Another part is in mass elements captured by the reference N-body simulation. As in Section~\ref{sec:sims}, the distribution of this mass may be estimated by examining regions of the simulation with similar distributions of visible mass, although in this case all the particles in the simulation may be used rather than just well-resolved halos because there is no danger of double-counting mass. This provides a plausible small-scale configuration of the smoothed mass (and hence the corresponding $\Phi$, $a$ and $K$ at the test point), and the possible configurations may then be marginalised over to estimate the uncertainty in the proxy values. Any remaining mass must be located in structures below the resolution limit of the simulation, and the maximum-entropy assumption for this mass is that it is smoothly distributed, as in the original field. This procedure has the further advantage of reconstructing modes of the power spectrum below the $2.4 \: h^{-1}$ Mpc limit of~\citet{Lavaux} field alone (see their figure 2), enhancing the utility of that framework.}

\item{} \red{It is necessary to add by hand the small-scale power due to halos because the 2M++ density field cannot be resolved below $2.4 \: h^{-1}$ Mpc. These halos come from a simulation box bearing no relation to the local universe, necessitating the correlation with observables described in Section~\ref{sec:sims}. However, the analysis of~\citet{Lavaux} also infers the initial conditions of structure formation in the 2M++ volume. This opens the possibility of running simulations constrained to match the smoothed density field at $z=0$. Such simulations would automatically place the correct mass in the simulation volume, as well as producing realisations of the density field resolved to the level of the particle mass. The distributions of $\Phi$, $a$ and $K$ maps would then be obtained by running multiple simulations consistent with the 2M++ number density and peculiar velocity fields, thereby marginalising over both the smoothed $z=0$ density field and the mass distribution on smaller scales. This method has the potential to obviate the need for all the inputs of Table~\ref{tab:params} beside the second.}

\end{enumerate}

\subsubsection{Further considerations for tests of gravity}

\red{A range of possible $\Phi$, $\vec{a}$ and $K$ fields corresponds to an uncertainty in the degree of screening of a given galaxy (under scalar--tensor gravity) or in its effective force law (under MOND). This will introduce marginal cases when dividing a test galaxy sample into different gravitational regimes, weakening null tests of GR and constraints on modified gravity parameters. The magnitude of the total uncertainty therefore determines the combination of minimum signal strength and sample size that would be statistically significant.}

A full comparison with modified gravity expectations would have further uncertainties. To make predictions for a specific theory, it would likely be invalid in detail to assume $\Lambda$CDM to calculate the halo mass function, the galaxy--halo connection, the relation between absolute and apparent magnitude, or the quasi-linear modes of the 2M++ field. We justify our doing so by arguing that the types of modified gravity theory that motivate our work are likely to generate larger deviations from GR on galactic than cosmological scales. The impact on cosmology -- and hence those parts of our analysis that rely on $\Lambda$CDM -- would be a second order effect in the prediction of the final signals.

This argument is \textit{a priori} least plausible in the case of MOND, which is unique among the models motivating our work in that it seeks to eliminate dark matter. MOND is not connected continuously to GR and $\Lambda$CDM as scalar-tensor theories are, but rather postulates a discrete deviation at the level of the Newtonian limit. The level of systematic error between our calculated $\vec{a}$ and that in MOND likely depends on the specific formulation of the theory. In QUMOND~\citep{QUMOND}, the MOND acceleration is simply an algebraic function of the Newtonian acceleration due to baryons alone, and the difference may therefore be represented by a distribution of ``phantom dark matter.'' Although there is no \textit{a priori} reason why this distribution should be correctly given by AM (for one there is no longer a halo mass function), the fact that observational estimates of the relation between visible and dynamical mass roughly agree with the AM prediction~\citep{Behroozi} suggests that this technique would give a sensible estimate for the amount of phantom dark matter as a function of luminosity in the mass range of interest. We would however expect $c_a$ to be nearer 1 in this case, as there would be no contribution to $\vec{a}$ from mass unassociated with light. As this tends to increase $|\vec{a}|$, our result would constitute an upper limit on the external acceleration field in MOND, making our conclusion that there are very few objects in the fully external field-dominated regime robust.

Other formulations, however, include a curl field in the relation between Newtonian and MONDian acceleration, which introduces further nonlinearities into the dependence of the total acceleration on the (baryonic) mass distribution, in addition to a potential offset between the direction of the true acceleration and the one we have calculated (e.g.~\citealt{Llinares,Famaey_McGaugh}). In such cases our method may give a poor approximation to the acceleration governing the MOND external field effect; we leave further investigation of this effect to future work.

\section{Conclusion}
\label{sec:conclusion}

We have combined an all-sky galaxy catalogue, a high-resolution N-body box and an estimate of the smoothed density field of the local universe to map the gravitational potential $\Phi$, acceleration $\vec{a}$ and curvature $K$ to a distance of 200 Mpc. As the fundamental variables of the gravitational field, these may be expected on general grounds to mark the transitions between gravitational regimes, and at the galaxy scale they are of specific importance in screened scalar-tensor theories and EP-violating models such as MOND. We use our maps to determine the sensitivity that galaxy-scale tests of modified gravity may be expected to achieve, and identify promising regions of the local universe. Our method may be easily adapted to any source and test galaxy sample, and we make public the code for doing so.

From the overall distributions of the gravitational variables we draw four general conclusions.

\begin{itemize}

\item{} When calculated in spheres of radius 10 (3) Mpc (typical Compton wavelengths of a light scalar field), the distribution of $\Phi/c^2$ peaks at around $10^{-5}$ ($10^{-6}$), and has a minimum of $10^{-6}$ (few $\times 10^{-8}$) set largely by the smooth component of the density field. Since the background scalar field in chameleon-screened scalar-tensor theories (e.g. $f_{R0}$ in $f(R)$ gravity) is of order the threshold in $\Phi/c^2$ at which screening becomes operative, astrophysical measurements are capable of constraining this to at least the $10^{-7}$ level by investigating galaxies with low internal $\Phi$ values. This makes galaxy-scale tests of screening significantly stronger than cosmological probes (e.g.~\citealt{Lombrisier}).

\item{} The external acceleration field is typically around a few $\times 10^{-15} \: \text{km} \, \text{s}^{-2}$ and rarely exceeds the characteristic acceleration $a_0 \approx 10^{-13} \: \text{km} \, \text{s}^{-2}$ at which visible and dynamical masses diverge. Thus, in the context of MOND-type models, few objects will generically be in the fully external field-dominated regime in which dynamics is Newtonian even for arbitrarily small internal acceleration. Locating such objects to test the external field effect would therefore require targeted searches around known mass concentrations.

\item{} Curvature values are centred around $10^{-55} \: \text{cm}^{-2}$, at the low end of the ``curvature desert'' that~\citet{Baker} identifies as a potentially interesting region for modified gravity.

\item{} The precision with which $\Phi$, $\vec{a}$ and $K$ may be mapped is limited by our ability to detect faint objects at large distance, and galaxy catalogues from future surveys such as DESI and LSST will therefore greatly reduce uncertainties. Theoretical advances such as the creation of simulations constrained to match the local volume may also be used to improve precision.

\end{itemize}

\red{In future work we will improve our total mass estimation, model fully the uncertainties in our maps by Monte Carlo marginalisation, and apply our framework to specific tests of gravity in galaxies.}

\section*{Acknowledgements}

We thank Julien Devriendt, Bhuvnesh Jain, Vinu Vikram and Risa Wechsler for helpful discussions during the course of this work, and Tessa Baker, Gary Mamon, Stacy McGaugh and an anonymous referee for comments on the manuscript. HD received support from the Balzan foundation via New College, Oxford, the U.S.\ Department of Energy under contract number DE-AC02-76SF00515, and St John's College, Oxford. PGF is supported by ERC H2020 693024 GravityLS project, the Beecroft Trust and STFC. GL acknowledges financial support from ``Programme National de Cosmologie and Galaxies'' (PNCG) of CNRS/INSU, France. GL is supported by ANR grant ANR-16-CE23-0002.

This work made use of the Dark Sky simulations, which were produced using an INCITE 2014 allocation on the Oak Ridge Leadership Computing Facility at Oak Ridge National Laboratory. We thank the Dark Sky collaboration for creating and providing access to this simulation,
and Sam Skillman and Yao-Yuan Mao for running \textsc{rockstar} and \textsc{consistent trees} on it, respectively. Additional computation was performed at SLAC National Accelerator Laboratory.

This research was supported by the DFG cluster of excellence ``Origin and Structure of the Universe'' (\url{www.universe-cluster.de}).
This work was partly made in the ILP LABEX (under reference ANR-10-LABX-63) supported by French state funds managed by the ANR within the Investissements d'Avenir programme under reference ANR-11-IDEX-0004-02.

\bsp


\begin{thebibliography}{10}

\bibitem[Babichev et al.(2009)]{kinetic}
Babichev E., Deffayet C., Ziour R., 2009, International Journal of Modern Physics D, 18, 2147 

\bibitem[Baker et al.(2013)]{PPF}
Baker T., Ferreira P.~G., Skordis C., 2013, PRD, 87, 024015 

\bibitem[Baker et al.(2015)]{Baker} Baker T., Psaltis D., Skordis C., 2015, ApJ, 802, 63 

\bibitem[Behroozi et al.(2010)]{Behroozi}
Behroozi P.~S., Conroy C., Wechsler R.~H., 2010, ApJ, 717, 379 

\bibitem[\protect\citeauthoryear{Behroozi, Wechsler \& Wu}{Behroozi et al.}{2013}]{Rockstar_1}
Behroozi P.~S., Wechsler R.~H., Wu H.-Y., 2013, ApJ, 762, 109 

\bibitem[Bell et al.(2003)]{Bell}
Bell E.~F., McIntosh D.~H., Katz N., Weinberg M.~D., 2003, ApJS, 149, 289 

\bibitem[\protect\citeauthoryear{Blumenthal et al.}{1986}]{Blumenthal}
Blumenthal G.~R., Faber S.~M., Flores R., Primack J.~R., 1986, ApJ, 301, 27

\bibitem[Bull et al.(2016)]{Bull}
Bull P. et al., 2016, Physics of the Dark Universe, 12, 56 

\bibitem[Cabr{\'e} et al.(2012)]{Cabre}
Cabr{\'e} A., Vikram V., Zhao G.-B., Jain B., Koyama K., 2012, JCAP, 7, 034 

\bibitem[Clifton et al.(2012)]{ModGrav}
Clifton T., Ferreira P.~G., Padilla A., Skordis C., 2012, PhysRep, 513, 1 

\bibitem[Cole \& Lacey(1996)]{Cole_Lacey}
Cole S., Lacey C., 1996, MNRAS, 281, 716 

\bibitem[\protect\citeauthoryear{Conroy, Wechsler \& Kravtsov}{Conroy et al.}{2006}]{Conroy}
Conroy C., Wechsler R.~H., Kravtsov A.~V., 2006, ApJ, 647, 201 

\bibitem[Desmond(2017)]{D16}
Desmond H., 2017, MNRAS, 464, 4160 

\bibitem[Desmond \& Wechsler(2015)]{DW15}
Desmond H., Wechsler R.~H., 2015, MNRAS, 454, 322 

\bibitem[Desmond \& Wechsler(2016)]{DW16}
Desmond H., Wechsler R.~H., 2017, MNRAS, 465, 820

\bibitem[Diemer \& Kravtsov(2015)]{Diemer_Kravtsov}
Diemer B., Kravtsov A.~V., 2015, ApJ, 799, 108

\bibitem[Famaey \& McGaugh(2012)]{Famaey_McGaugh}
Famaey B., McGaugh S.~S., 2012, Living Rev. Relativ., 15, 10 

\bibitem[Governato et al.(2010)]{Governato}
Governato F. et al., 2010, Nature, 463, 203 

\bibitem[Hu \& Sawicki(2007)]{Hu_Sawicki}
Hu W., Sawicki I., 2007, PRD, 76, 064004 

\bibitem[Hui et al.(2009)]{Hui}
Hui L., Nicolis A., Stubbs C.~W., 2009, PRD, 80, 104002 

\bibitem[Jain(2011)]{Jain_Surveys}
Jain B., 2011, Philosophical Transactions of the Royal Society of London Series A, 369, 5081 

\bibitem[Jain \& Khoury(2010)]{Jain_Khoury}
Jain B., Khoury J., 2010, Annals of Physics, 325, 1479 

\bibitem[Jain \& VanderPlas(2011)]{Jain_Vanderplas}
Jain B., VanderPlas J., 2011, JCAP, 10, 032 

\bibitem[Jain et al.(2013)]{Jain_NovelProbes}
Jain B. et al., 2013, preprint (arXiv:1309.5389)

\bibitem[Khoury \& Weltman(2004)]{Chameleon}
Khoury J., Weltman A., 2004, PRD, 69, 044026 

\bibitem[Khoury(2013)]{Khoury_LesHouches}
Khoury J., 2013, preprint (arXiv:1312.2006) 

\bibitem[Kravtsov et al.(2004)]{Kravtsov}
Kravtsov A.~V., Berlind A.~A., Wechsler R.~H., Klypin A.~A., Gottlober S., Allgood B., Primack J.~R., 2004, ApJ, 609, 35 

\bibitem[Lavaux \& Hudson(2011)]{2M++}
Lavaux G., Hudson M.~J., 2011, MNRAS, 416, 2840 

\bibitem[Lavaux \& Jasche(2016)]{Lavaux}
Lavaux G., Jasche J., 2016, MNRAS, 455, 3169 

\bibitem[Lehmann et al.(2015)]{Lehmann}
Lehmann B.~V., Mao Y.-Y., Becker M.~R., Skillman S.~W., Wechsler R.~H., 2015, preprint (arXiv:1510.05651)

\bibitem[Lelli et al.(2017)]{Lelli}
Lelli F., McGaugh S.~S., Schombert J.~M., Pawlowski M.~S., 2017, ApJ, 836, 152 

\bibitem[Llinares et al.(2008)]{Llinares}
Llinares C., Knebe A., Zhao H., 2008, MNRAS, 391, 1778 

\bibitem[Lombriser et al.(2012)]{Lombrisier}
Lombriser L., Slosar A., Seljak U., Hu W., 2012, PRD, 85, 124038 

\bibitem[McGaugh(1999)]{MG99}
McGaugh S.~S., 1999, Galaxy Dynamics -- A Rutgers Symposium, Astron. Soc. Pacific Conf. Ser., ed. D.R. Merritt, M. Valluri, J.A. Sellwood, Vol. 182 (San Francisco: ASP), p. 528

\bibitem[McGaugh \& Milgrom(2013a)]{McGaugh_EFE}
McGaugh S., Milgrom M., 2013, ApJ, 766, 22 

\bibitem[McGaugh \& Milgrom(2013b)]{McGaugh_EFE_2}
McGaugh S., Milgrom M., 2013, ApJ, 775, 139 

\bibitem[Milgrom(1983a)]{Milgrom1}
Milgrom M., 1983, ApJ, 270, 365

\bibitem[Milgrom(1983b)]{Milgrom2}
Milgrom M., 1983, ApJ, 270, 371 

\bibitem[Milgrom(1983c)]{Milgrom3}
Milgrom M., 1983, ApJ, 270, 384

\bibitem[Milgrom(2010)]{QUMOND}
Milgrom M., 2010, MNRAS, 403, 886 

\bibitem[Moster et al.(2010)]{Moster}
Moster B.~P., Somerville R.~S., Maulbetsch C., van den Bosch F.~C., Macci{\`o} A.~V., Naab T., Oser L., 2010, ApJ, 710, 903

\bibitem[Navarro et al.(1996)]{NFW}
Navarro J.~F., Frenk C.~S., White S.~D.~M., 1996, ApJ, 462, 563 

\bibitem[Nordtvedt(1968)]{Nordtvedt}
Nordtvedt K., 1968, Physical Review, 169, 1017 

\bibitem[Pontzen \& Governato(2012)]{Pontzen}
Pontzen A., Governato F., 2012, MNRAS, 421, 3464 

\bibitem[Reddick et al.(2013)]{Reddick}
Reddick R.~M., Wechsler R.~H., Tinker J.~L., Behroozi P.~S., 2013, ApJ, 771, 30

\bibitem[Reed et al.(2013)]{Reed}
Reed D.~S., Smith R.~E., Potter D., Schneider A., Stadel J., Moore B., 2013, MNRAS, 431, 1866 

\bibitem[Schlamminger et al.(2008)]{EP}
Schlamminger S., Choi K.-Y., Wagner T.~A., Gundlach J.~H., Adelberger E.~G., 2008, PRL, 100, 041101

\bibitem[Schlegel et al.(1998)]{Schlegel}
Schlegel D.~J., Finkbeiner D.~P., Davis M., 1998, ApJ, 500, 525 

\bibitem[Skillman et al.(2014)]{DarkSky}
Skillman S.~W., Warren M.~S., Turk M.~J., Wechsler R.~H., Holz D.~E., Sutter P.~M., 2014, preprint (arXiv:1407.2600)

\bibitem[Trujillo-Gomez et al.(2011)]{TG}
Trujillo-Gomez S., Klypin A., Primack J., Romanowsky A.~J., 2011, ApJ, 742, 16 

\bibitem[Vainshtein(1972)]{Vainshtein}
Vainshtein A.~I., 1972, Phys. Lett. B, 39, 393 

\bibitem[van Daalen \& Schaye(2015)]{vanDaalen}
van Daalen M.~P., Schaye J., 2015, MNRAS, 452, 2247 

\bibitem[Vikram et al.(2013)]{Vikram}
Vikram V., Cabr{\'e} A., Jain B., VanderPlas J.~T., 2013, JCAP, 8, 020 

\bibitem[Wang et al.(2012)]{Wang}
Wang J., Hui L., Khoury J., 2012, PRL, 109, 241301 

\bibitem[Warren(2013)]{Warren13}
Warren M.~S., 2013, Proc. Int. Conf. High Perform. Comput. Netw. Storage Anal., (New York: ACM), p. 72

\bibitem[Will(1993)]{Will}
Will C.~M., 1993, Theory and Experiment in Gravitational Physics, Cambridge, CUP, 1993 

\end{thebibliography}
\end{document}